\documentclass[conference,a4paper,onecolumn]{IEEEtran}

\usepackage{amsthm}
\usepackage{amssymb}
\usepackage{epsfig}
\usepackage{epstopdf}
\usepackage{graphicx}
\usepackage{graphics}
\usepackage{cite}
\usepackage{mathrsfs}
\usepackage{dsfont}
\usepackage{euscript}
\usepackage[utf8]{inputenc}
\usepackage[T1]{fontenc}
\usepackage{url}
\usepackage{ifthen}
\usepackage{subcaption}
\usepackage[cmex10]{amsmath} 

\newtheorem{thm}{Theorem}

\newtheorem{prop}[thm]{Proposition}
\theoremstyle{definition}

\theoremstyle{remark}
\newtheorem{rem}[thm]{Remark}

\DeclareMathOperator{\IM}{Im}

\newcommand{\snr}{ {\mathsf{snr}} }

\newcommand{\abs}[1]{\left\vert#1\right\vert}
\newcommand{\set}[1]{\left\{#1\right\}}

\newcommand{\vct}[1]{\mathbf{#1}}
\newcommand{\Mat}[1]{\mathbf{#1}}

\newcommand\eg{{{e.g.}}}

\newcommand{\comment}[1]  {}
\def\BE{\begin{equation}}
\def\EE{\end{equation}}
\def\BEA{\begin{eqnarray}}
\def\EEA{\end{eqnarray}}

\newcommand\mA{{\bf A}} 

\newcommand{\bbC}{\mathbb{C}}
\newcommand{\bbE}{\mathbb{E}}
\newcommand{\bbN}{\mathbb{N}}

\newcommand{\bbR}{\mathbb{R}}

\newcommand{\calA}{\mathcal{A}}

\newcommand{\calG}{\mathcal{G}}
\newcommand{\calH}{\mathcal{H}}

\newcommand{\calL}{\mathcal{L}}

\newcommand{\calP}{\mathcal{P}}

\newcommand{\calS}{\mathcal{S}}

\newcommand{\rmI}{\mathrm{I}}

\newcommand{\sfR}{\mathsf{R}}

\newcommand{\sfm}{\mathsf{m}}

\newcommand{\EuS}[1]{\EuScript{#1}}

\newcommand{\rmd}{\mathrm{d}}

\newcommand{\sj}{\mathsf{j}}

\newcommand{\opt}{\mathrm{opt}}
\newcommand{\ms}{\mathrm{mmse}}

\newcommand{\febno}{\frac{E_b}{N_0}}

\usepackage[usenames,dvipsnames,svgnames,table]{xcolor}

\usepackage[colorlinks=true,citecolor=Blue,linkcolor=BrickRed]{hyperref}

\makeatletter
\renewcommand*{\eqref}[1]{%
  \hyperref[{#1}]{\textup{\tagform@{\ref*{#1}}}}%
}
\makeatother

\begin{document}
\title{Sparse NOMA: A Closed-Form Characterization}

\author{\IEEEauthorblockN{Benjamin M. Zaidel
}
\IEEEauthorblockA{Faculty of Engineering\\
Bar-Ilan University\\
Ramat-Gan 52900, Israel\\
Email: benjamin.zaidel@gmail.com}
\and
\IEEEauthorblockN{Ori Shental
}
\IEEEauthorblockA{Communications Theory Department\\
Bell Labs   \\
Holmdel, New Jersey 07733, USA\\
Email: ori.shental@nokia-bell-labs.com}
\and
\IEEEauthorblockN{Shlomo Shamai (Shitz)
}
\IEEEauthorblockA{Department of Electrical Engineering\\
Technion\\ Haifa 32000, Israel\\
Email: sshlomo@ee.technion.ac.il
}}

\maketitle

\begin{abstract}
Understanding fundamental limits of the various technologies suggested for future
5G and beyond
cellular systems is crucial for developing efficient state-of-the-art designs.
A leading technology of major interest is non-orthogonal multiple-access (NOMA). In this paper, we derive an explicit rigorous \emph{closed-form} analytical expression for the optimum spectral efficiency in the large-system limit of \emph{regular} sparse NOMA, where only a \emph{fixed and finite} number of orthogonal resources are allocated to any designated user, and vice versa.
The basic Verd\'{u}-Shamai
formula for (dense) randomly-spread code-division multiple-access (RS-CDMA)
turns out to coincide with the limit of the derived expression, when the number of orthogonal resources per user grows large.
Furthermore, regular sparse NOMA is rigorously shown to be spectrally more efficient than
RS-CDMA
across the entire system load range.
It may therefore serve as an efficient means for reducing the throughput gap to orthogonal transmission in the underloaded regime, and to the ultimate Cover-Wyner bound in  overloaded systems.
The results analytically reinforce preliminary conclusions in  \cite{Shental-Zaidel-Shamai-ISIT-2017}, which mostly relied on heuristics and numerical observations.
The spectral efficiency is also derived in closed form
for the sub-optimal linear minimum-mean-square-error (LMMSE) receiver, which again extends the corresponding Verd\'{u}-Shamai LMMSE formula to regular sparse NOMA.
\end{abstract}


\section{Introduction}
Orthogonal transmission is the holy grail of multiple-access communications. Orthogonal multiple-access (OMA) allows for a feasible transceiver design, which also achieves the ultimate total throughput in a fully-loaded Gaussian channel, where the number of users equals the total number of available orthogonal resources.
However, non-orthogonal multiple-access (NOMA) schemes with tractable near-optimal receivers open up the practicality of operation in the overloaded regime, where the number of designated users \emph{exceeds} the number of available resources.
Note that even for pragmatic underloaded systems it is typically hard to maintain orthogonality, and OMA leads to severe throughput penalties. Thus, NOMA exhibits significant advantages over OMA by either supporting more concurrent users or, alternatively, facilitating higher user throughput when orthogonality breaks down. In a nutshell, these are the main incentives driving NOMA as a key enabler for the Internet-of-Things (IoT) and 5G and beyond networks. An overview of the ample literature on NOMA can be found, e.g., in \cite{Ding-Lei-Karagiannidis-Schober-Yuan-Bhargava-2017} and references therein.

Sparse, or low-density code-domain (LDCD) NOMA is a prominent sub-category, which conceptually relies on multiplexing low-density signatures (LDS) \cite{Hoshyar-Razavi-Al-Imari-VTC-2010}. Sparse spreading codes comprising only a small number of non-zero elements are employed for linearly modulating each user's symbols over shared physical orthogonal resources. The sparse mapping between users and resources is dubbed: \emph{regular} when each user occupies a fixed number of resources, and each resource is used by a fixed number of users; \emph{irregular} when the respective numbers are random, and only fixed on average; \emph{partly-regular} when each user occupies a fixed number of resources, and each resource is used by a random, yet fixed on average, number of users (or vice versa).
The main attractiveness of this central class of NOMA schemes is in its inherent receiver complexity reduction, achieved by utilizing message-passing algorithms (MPAs)
\cite{Ding-Lei-Karagiannidis-Schober-Yuan-Bhargava-2017}. Different variants of sparse NOMA, \eg, sparse-code multiple-access (SCMA) \cite{Nikopour-Baligh-PIMRC-2013}, have recently gained much attention in 5G standardization.
Naturally, information-theoretic analysis of sparse NOMA
has been a fruitful grounds for research,
 providing a great deal of insight into its workings, e.g.,~\cite{Yoshida-Tanaka-ISIT-2006,
 Guo-Wang-2008,Guo-Baron-Shamai-Allerton-2009,
 Ferrante-Di-Benedetto-2015}.
 However, it is important to note that neither of these contributions provide an explicit rigorous closed-form analytical characterization
of sparse NOMA throughput. This holds even in the large-system limit, and  comes in sheer contrast to the insightful work by Verd\'{u} and Shamai on
(dense)
randomly-spread code-division multiple-access (RS-CDMA) \cite{Verdu-Shamai-paper-3-99}.

In a recent paper~\cite{Shental-Zaidel-Shamai-ISIT-2017}, the authors have presented a closed-form analytical expression for the \emph{limiting empirical squared singular value density} of a spreading (signature) matrix, corresponding to \emph{regular} sparse NOMA.
The result was rigorously derived only for a repetition-based sparse spreading scheme, while for binary sparse random spreading the same result was claimed to hold based on the \emph{heuristic} cavity method of statistical physics.
The limiting density was then used to plot the optimum asymptotic spectral efficiency, but only by means of numerical integration.

In this contribution, using a recent fundamental result by Bordenave and Lelarge \cite{Bordenave-Lelarge-2010},
we first re-establish the aforementioned limiting density result in full rigor, and in fact extend it to the more general case of sparse signatures, of which non-zero entries reside on the unit-circle in the complex plane.
Then,
a rigorous \emph{closed-form explicit analytical expression} for the spectral efficiency of regular sparse NOMA with optimum decoding is derived. The formula is obtained for Gaussian signaling and non-fading channels in the asymptotic large-system limit.  The Verd\'{u}-Shamai optimum spectral efficiency formula~\cite{Verdu-Shamai-paper-3-99} turns out to be a particular instantiation of the derived expression for the case where the spreading codes become dense.
A closed-form expression for the asymptotic spectral efficiency of the linear minimum-mean-square-error (LMMSE) receiver is also provided. An extreme-SNR characterization leading to useful insights is given as well.
The results allow for easy comparison to the spectral efficiency of
RS-CDMA \cite{Verdu-Shamai-paper-3-99} (representing \emph{dense} NOMA),
and other sparse NOMA variants, e.g., \cite{Yoshida-Tanaka-ISIT-2006,Ferrante-Di-Benedetto-2015}.
Assuming optimal decoding, the remarkable superiority of the feasible regular sparse NOMA over not only its irregular, or partly-regular, counterparts, but also over the intractable RS-CDMA, as originally observed (only) numerically in~\cite{Shental-Zaidel-Shamai-ISIT-2017}, is here being analytically verified.



\section{Sparse NOMA System Model}\label{sec: Sparse NOMA System Model}

Consider a system where the signals of $K$ users are multiplexed over $N$ shared orthogonal resources. These resources can designate, e.g., orthogonal frequencies in orthogonal frequency-division multiplexing (OFDM) based systems, or different time-slots in time-hopping (TH)
multiple-access systems~\cite{Ferrante-Di-Benedetto-2015}. Multiplexing is performed while employing randomly chosen sparse spreading signatures of length $N$ (namely, $N$-dimensional vectors). Each sparse signature is assumed to contain a small number of non-zero entries (typically much smaller than $N$), while the remaining entries are set to zero.
The $N$-dimensional received signal, at some arbitrary time instance, at the output of a generic complex Gaussian vector channel, adheres to
\begin{IEEEeqnarray}{rCl}\label{eq: System model}
    \vct{y}&=&\sqrt{\tfrac{\snr}{d}}\Mat{A}\vct{x}+\vct{n} \ ,
\end{IEEEeqnarray}
where $\vct{x}$ is a $K$-dimensional complex vector comprising the coded symbols of the users. Assuming Gaussian signaling, full symmetry, fixed powers, and no cooperation between encoders corresponding to different users, the input vector $\vct{x}$ is distributed as
$\vct{x}\sim\mathcal{CN}(\vct{0},\Mat{I}_{K})$;
i.e., a unit energy per symbol per user is assumed.
$\Mat{A}$ denotes the $N\times K$ sparse signature matrix, whose $k$th column represents the spreading signature of user $k$, and its non-zero entries designate the user-resource mapping (user $k$ occupies resource $n$ if $A_{nk}\neq0$, where $A_{nk}$ denotes the $(n,k)$'th entry of $\Mat{A}$). Finally,
$\vct{n}\sim\mathcal{CN}(\mathbf{0},\Mat{I}_{N})$
denotes the $N$-dimensional circularly-symmetric complex additive white Gaussian noise (AWGN) vector at the receiving end. The normalization factor $\frac{1}{\sqrt{d}}$ in \eqref{eq: System model}, $d\in\mathbb{N}^+$,
controls the signature norms (as explained in the sequel).
The parameter $\snr$ thus designates the received signal-to-noise ratio (SNR) of each user, and we further use  $\beta\triangleq \frac{K}{N}$ to denote the \emph{system load}~(users per resource).

In sparse NOMA, due to the sparsity of $\Mat{A}$, typically only a few of the users' signals collide over any given orthogonal resource. The regularity assumption dictates that each column of $\mA$ (respectively, row) has \emph{exactly} $d$ (respectively, $\beta d$) non-zero entries.
$\beta$ is therefore chosen here so that $\beta d\in\mathbb{N}^{+}$.
The non-zero entries of $\mA$ are assumed to arbitrarily reside on the unit-circle in the complex plane.
Repetition-based spreading and random binary spreading \cite{Shental-Zaidel-Shamai-ISIT-2017} thus constitute special cases.
The model may also account for phase-fading scenarios in conjunction with sparse spreading.
The normalization in~\eqref{eq: System model} ensures that the columns of $\frac{1}{\sqrt{d}}{\Mat{A}}$ have unit norm.
Finally, we also assume here that  $\Mat{A}$ is perfectly known at the receiving end, and uniformly chosen randomly and independently per each channel use from the set of
$(\beta d,d)$-regular matrices.

Note at this point that a key tool in the derivations to follow is the observation that the signature matrix $\Mat{A}$ can be associated with the \emph{adjacency matrix} of a random $(\beta d,d)$-semiregular bipartite (factor) graph $\mathcal{A}$, where a user node $k$ and a resource node $n$ are connected if and only if $A_{nk}\neq0$. This factor graph is assumed henceforth to be \emph{locally tree-like}\footnote{A  precise mathematical definition can be found, e.g., in \cite{Bordenave-Lelarge-2010}.},
and to converge in the large-system limit (as $N\to\infty$) to a \emph{bipartite Galton-Watson tree} (BGWT), as specified in Section \ref{sec: Asymptotic Spectral Density}.
This assumption essentially implies that for large dimensions
short cycles are rare (similar to LDPC codes \cite{Bordenave-Lelarge-2010}), and it allows for reduced complexity iterative near-optimal multiuser detection by applying MPAs over the underlying factor graph.

\section{Asymptotic Spectral Density}\label{sec: Asymptotic Spectral Density}

Our first step is to characterize the limiting empirical distribution of the squared singular values  of the normalized signature matrix $\frac{1}{\sqrt{d}}\Mat{A}$, which is a cornerstone in the derivation of our main analytical results.
To this end, we first review a useful result by Bordenave and Lelarge \cite{Bordenave-Lelarge-2010} on  properties of random weighted bipartite graphs, whose random weak limit is the probability measure of a BGWT.
 Due to space limitations, we avoid a fully formal representation, while referring the reader to \cite{Bordenave-Lelarge-2010} for a full account.

Consider a sequence of random bipartite graphs $\set{\calG_N}$, converging in law to a BGWT with degree distribution $(\delta_{\beta d},\delta_d)$, and parameter $\frac{1}{1+\beta}$. Let $\Mat{W}$ be an
$N\times K$
complex random matrix independent of $\calG_N$,
with i.i.d.\  entries having finite absolute second moments.
With some abuse of notation, let the weighted adjacency matrix of $\calG_N$ read
\begin{IEEEeqnarray}{rCL}\label{eq: Adjacency matrix of a weighted bipartite graph}
    \tilde{\Mat{A}}_N &=& \begin{pmatrix}
                \Mat{0} & \Mat{A} \\
                \Mat{A}^\dag & \Mat{0} \\
              \end{pmatrix} \triangleq
              \begin{pmatrix}
                \Mat{0} & \Mat{W} \\
                \Mat{W}^\dag & \Mat{0} \\
              \end{pmatrix}
               \circ
  \begin{pmatrix}
                \Mat{0} & \bar{\Mat{A}} \\
                \bar{\Mat{A}}^\dag & \Mat{0} \\
              \end{pmatrix}  \ ,
\end{IEEEeqnarray}
where $\circ$ denotes the Hadamard product, $\bar{\Mat{A}}$ is an $N\times K$ matrix such that $\bar{{A}}_{ij}=1$ if vertex $i$ is connected by an edge to vertex $N+j$ on $\calG_N$, and $\bar{{A}}_{ij}=0$ otherwise, where $i\in\set{1,\dots,N}$, and $j\in\set{1,\dots,K}$.
Let $\calH$ denote the set of holomorphic functions $f:\bbC^+\to \bbC^+$ such that $\abs{f(z)}\le \frac{1}{\IM(z)}$. Finally, let $\calP(\calH)$ denote the space of probability measures on $\calH$.
\begin{thm}[\!\!{\cite[Theorems 4 \& 5]{Bordenave-Lelarge-2010}}]\label{thm: Bordenave and Lelarge Theorems 4&5}
Let the above assumptions on the structure of $ \tilde{\Mat{A}}_N$ hold. Then
\begin{enumerate}
\item\label{thm: Part I of BL Theorem} There exists a unique pair of probability measures $(\mu_a,\mu_b)\in \calP(\calH)\times \calP(\calH)$ such that
for all $z\in\bbC^+$
\begin{IEEEeqnarray}{rCl}
Y^a(z) &\overset{d}{=}& - (z+\textstyle{\sum}_{i=1}^{\beta d -1} \abs{W^b_i}^2 Y_i^b(z))^{-1}
\label{eq: RDE for Ya}\\
Y^b(z) &\overset{d}{=}& - (z+\textstyle{\sum}_{i=1}^{d -1}\abs{W^a_i}^2 Y_i^a(z))^{-1} \ ,
\label{eq: RDE for Yb}
\end{IEEEeqnarray}
where $Y^a,Y_i^a$ (respectively, $Y^b,Y_i^b$) are i.i.d.\ random variables with law $\mu_a$ (respectively, $\mu_b$), and $W^a_i,W^b_i$ are i.i.d. random variables distributed as $W_{11}$\footnote{The equality $\overset{d}{=}$ in \eqref{eq: RDE for Ya} and \eqref{eq: RDE for Yb} is in the sense that the \emph{distribution} of the random variables on both sides of the equation is the same.}.
\item For all $z\in\bbC^+$, the Stieltjes transform\footnote{The Stieltjes transform of a probability measure $\mu$ on $\bbR$ reads $m(z)\triangleq\int_{\bbR} \frac{1}{x-z}\,\rmd \mu(x)$, $z\in\bbC^+$. The measure $\mu$ can be recovered from $m(z)$ via the Stieltjes inversion formula $\rmd \mu(\lambda) = \frac{1}{\pi} \lim_{\epsilon\to 0^+} \IM (m(z))\vert_{z=\lambda+\sj \epsilon} \, \rmd \lambda$, where the limit is in the sense of weak convergence of measures (e.g., \cite{Tulino-Verdu-2004}).} $m_{\tilde{\Mat{A}}_N}(z)$ of the empirical eigenvalue distribution of $\tilde{\Mat{A}}_N$ converges as $N\to\infty$ in $L^1$ to $m_{\tilde{\Mat{A}}}(z)=\frac{1}{1+\beta}\bbE\{X^a(z)\} + \frac{\beta}{1+\beta}\bbE\{X^b(z)\}$, where for all $z\in\bbC^+$
\begin{IEEEeqnarray}{rCl}
X^a(z) &\overset{d}{=}& - (z+\textstyle{\sum}_{i=1}^{\beta d} \abs{W^b_i}^2 Y_i^b(z))^{-1}
\label{eq: RDE for Xa}\\
X^b(z) &\overset{d}{=}& - (z+\textstyle{\sum}_{i=1}^{d}\abs{W^a_i}^2 Y_i^a(z))^{-1} \ ,
\label{eq: RDE for Xb}
\end{IEEEeqnarray}
where $Y_i^a,Y_i^b,W^a_i$, and $W^b_i$ are i.i.d.\ copies with laws as in Part (\ref{thm: Part I of BL Theorem}).
\end{enumerate}
\end{thm}

Turning to the sparse regular NOMA setting specified in Section \ref{sec: Sparse NOMA System Model}, we immediately observe that the graph $\calA$ associated with the signature matrix $\Mat{A}$ (cf.\ \eqref{eq: System model}) falls exactly within the framework of Theorem \ref{thm: Bordenave and Lelarge Theorems 4&5}, and its weighted adjacency matrix can be expressed as in \eqref{eq: Adjacency matrix of a weighted bipartite graph}. Furthermore, the underlying assumption on the spreading signatures dictates that all entries of the corresponding weight matrix $\Mat{W}$ have surely a unit absolute value. We thus get the following result.

\begin{thm}\label{thm: Asymptotic Stieltjes transform and spectral density}
Let $\Mat{A}$ be a sparse random $N\times K$ matrix with exactly $2\le d\in\mathbb{N}^{+}<\infty$ (respectively, $2\le\beta d\in\mathbb{N}^{+}<\infty$) non-zero entries in each column (respectively, row),  arbitrarily distributed over the unit-circle in $\bbC$. Assume that the $(\beta d,d)$-semiregular bipartite graph $\calA$ associated with $\Mat{A}$ is locally tree-like, with a BGWT having degree distribution $(\delta_{\beta d},\delta_d)$ and parameter $\frac{1}{1+\beta}$ as a weak limit. Let $\alpha\triangleq\frac{d-1}{d}$ and $\gamma\triangleq\frac{\beta d-1}{d}$. Then
\begin{enumerate}
  \item For all $z\in\mathbb{C}^{+}$, the Stieltjes transform of the empirical eigenvalue distribution of $\frac{1}{d}\Mat{A}\Mat{A}^{\dag}$ converges as $N\to\infty$ in $L^1$ to
\begin{IEEEeqnarray}{rCl}
m_{\frac{1}{d}\Mat{A}\Mat{A}^{\dag}}(z) 
&=& - \bigl(z-\tfrac{\beta}{1+\alpha\sfm(z)}\bigr)^{-1} \ ,
\label{eq: Limiting expression for the Stieltjes transform}
\end{IEEEeqnarray}
where $\sfm(z)$ solves the following \emph{deterministic} equation:
\begin{IEEEeqnarray}{rCl}
    \sfm(z) 
    &=& - \bigl(z-\tfrac{\gamma}{1+\alpha\sfm(z)}\bigr)^{-1} \ . \label{eq: Limiting expression for sf_m}
\end{IEEEeqnarray}

  \item Subject to the convergence of the Stieltjes transform, the weak limit of the empirical  eigenvalue distribution of $\frac{1}{d}\Mat{A}\Mat{A}^{\dag}$ as $N\to\infty$
  is a
  distribution with density
\begin{IEEEeqnarray}{rCl}
\rho(\lambda,\beta,d) &=& [1-\beta]^{+}\delta(\lambda)+\tfrac{\beta d}{2\pi} \tfrac{\sqrt{[\lambda-\lambda^-]^{+}[\lambda^+-\lambda]^{+}}}{\lambda(\beta d - \lambda)} \ , \quad \ \label{eq: Final equation for the spectral density in terms of lambda+-}
\end{IEEEeqnarray}
where $\lambda^{\pm}=(\sqrt{\alpha}\pm\sqrt{\gamma})^2$, $\delta(\lambda)$ is a unit point mass at $\lambda=0$, and $[z]^{+}\triangleq\max\{0,z\}$.
\end{enumerate}
\end{thm}
\begin{IEEEproof}[Proof Outline]
The weights $\{W_i^a,W_i^b\}$ in \eqref{eq: RDE for Ya} and~\eqref{eq: RDE for Yb} have  unit absolute values. The equations hence admit a unique \emph{deterministic} solution.  Next, note that the eigenvalues of
\begin{IEEEeqnarray}{rCl}
    \tilde{\Mat{A}}_N^2 = \begin{pmatrix}
                \Mat{A}\Mat{A}^{\dag} & \mathbf{0} \\
                \mathbf{0} & \Mat{A}^{\dag}\Mat{A}\\
              \end{pmatrix} \label{eq: tilde-A-sq}
\end{IEEEeqnarray}
are simply the eigenvalues of $\Mat{A}\Mat{A}^{\dag}$ together with those of $\Mat{A}^{\dag}\Mat{A}$ (which are in fact the same up to $\abs{K-N}$ additional zero eigenvalues). Furthermore, the limiting Stieltjes transform of the empirical eigenvalue distribution of $\tilde{\Mat{A}}_N^2$ admits the following relation $z m_{\tilde{\Mat{A}}^2}(z^2)=m_{\tilde{\Mat{A}}}(z)$. This lets us conclude that $m_{\frac{1}{d}\Mat{A}\Mat{A}^{\dag}}(z)=\sqrt{\frac{d}{{z}}}X^a(\sqrt{dz})$, where $X^a(z)$ is obtained from \eqref{eq: RDE for Xa}. Eqs.\ \eqref{eq: Limiting expression for the Stieltjes transform} and \eqref{eq: Limiting expression for sf_m} then follow after some algebra. Finally,
we get \eqref{eq: Final equation for the spectral density in terms of lambda+-} using the Stieltjes inversion formula.
\end{IEEEproof}

\section{Optimum Receiver}\label{sec: Optimum Receiver}

The fundamental figure of merit for system performance is taken here as the normalized  spectral efficiency (total throughput)  in bits/sec/Hz per dimension. For optimum processing this quantity corresponds to the ergodic sum-capacity, given by the normalized conditional input-output mutual information \cite{Verdu-Shamai-paper-3-99}:
\begin{IEEEeqnarray}{rCl}\label{eq: Sum capacity as normalized mutual information}
C_N^\opt(\snr,\beta,d) &\triangleq& \tfrac{1}{N} \rmI(\vct{x};\vct{y}|\Mat{A})
\IEEEnonumber \\ &=& \tfrac{1}{N}\,
\bbE\!\set{\log_2\det\left(\Mat{I}_N+\tfrac{\snr}{d}\Mat{A}\Mat{A}^\dag\right)} .\quad \ \
\end{IEEEeqnarray}
Focusing on the large-system limit, the asymptotic spectral efficiency of the optimum receiver corresponds to
\begin{IEEEeqnarray}{rCl}
C^\opt(\snr,\beta,d) &\triangleq& \lim_{N\to\infty}C^\opt_N(\snr,\beta,d) \ , \label{eq: Limiting total throughput}
\end{IEEEeqnarray}
where the existence of the limit follows from Theorem \ref{thm: Asymptotic Stieltjes transform and spectral density}. The following result is one of the main contributions of this paper.
\begin{thm}
\label{thm: Asymptotic spectral efficiency of the optimum receiver}
Let $d,\beta,\alpha$ and $\gamma$ be as in Theorem~\ref{thm: Asymptotic Stieltjes transform and spectral density}. Further
let $\tilde{\beta}\triangleq\frac{\alpha}{\gamma}$ and $\zeta\triangleq\frac{\beta d}{\gamma}$.
Then, the optimum spectral efficiency~\eqref{eq: Sum capacity as normalized mutual information} converges as $N\to\infty$ to
\begin{IEEEeqnarray*}{LR}
C^{\opt}(\snr,\beta,d) =
& \\ \phantom{C^{\opt}}
\tfrac{ \beta (d - 1) + 1}{2}\log_2\bigl(1+(\gamma+\alpha)\snr - \tfrac{1}{4}\EuS{F}(\gamma\snr,\tilde{\beta})\bigr)
&\\
\phantom{C^{\opt}}
+ (\beta-1) \log_2\bigl(1+\alpha\snr - \tfrac{1}{4}\EuS{F}(\gamma\snr,\tilde{\beta})\bigr) & \\
\phantom{C^{\opt}} - \tfrac{\beta(d-1)-1}{2} \log_2\left(\tfrac{(1+ \beta d\,\snr)^2}{\EuS{G}(\gamma\snr,\zeta,\tilde{\beta})}\right) \ , & \quad  \IEEEyesnumber \label{eq: Expression for C_opt with G-x-y-z}
\end{IEEEeqnarray*}
where (cf.\ \cite{Verdu-Shamai-paper-3-99})
$\EuS{F}(x,z)\triangleq(({x\left(1+\sqrt{z}\right)^2+1})^{1/2} -({x\left(1-\sqrt{z}\right)^2+1})^{1/2})^2$
and 
\begin{IEEEeqnarray*}{LR}\label{eq: Definition of the function G-x-y-z}
\EuS{G}(x,y,z) \triangleq & \\
\biggl(\tfrac{\sqrt{\left(y-(1-\sqrt{z})^2\right)\left(x(1+\sqrt{z})^2+1\right)}-\sqrt{\left(y-(1+\sqrt{z})^2\right)\left(x(1-\sqrt{z})^2+1\right)}}{\sqrt{y-(1-\sqrt{z})^2}-\sqrt{y-(1+\sqrt{z})^2}}\biggr)^2 & \\
\phantom{\EuS{G}} x,y,z\in\mathbb{R}^+, \ y\ge (1+\sqrt{z})^2 \ .  \IEEEyesnumber &
\end{IEEEeqnarray*}
\end{thm}
\begin{IEEEproof}[Proof Outline]
The space $\calH$ specified in Section \ref{sec: Asymptotic Spectral Density},
 equipped with an appropriate topology, is a complete separable metrizable compact space (see \cite{Bordenave-Lelarge-2010}). Considering the sequence of Stieltjes transforms
 converging to \eqref{eq: Limiting expression for the Stieltjes transform} by Theorem \ref{thm: Asymptotic Stieltjes transform and spectral density}, recall that
the \emph{ergodic} normalized sum-capacity \eqref{eq: Sum capacity as normalized mutual information}
is determined by the \emph{distribution} of the signature matrices. Therefore,
by the Skorokhod representation theorem (see, e.g., \cite[Theorem 7]{Bordenave-Lelarge-2010}), we can assume that the sequence of Stieltjes transforms and its limit (which is in fact a deterministic function) are  defined on a common probability space, and that the convergence to the limit is in the almost sure sense.
We next observe that by Hadamard's inequality, along with the unit absolute value of the non-zero entries of $\Mat{A}$, $\frac{1}{N} \log_2\det\left(\Mat{I}_N+\tfrac{\snr}{d}\Mat{A}\Mat{A}^\dag\right)\le \log_2(1+\beta\snr)<\infty$. This in turn implies that the sequence $\frac{1}{N} \log_2\det\left(\Mat{I}_N+\tfrac{\snr}{d}\Mat{A}\Mat{A}^\dag\right)$ is uniformly integrable. By the weak convergence stated in Theorem \ref{thm: Asymptotic Stieltjes transform and spectral density} (cf.\ \eqref{eq: Final equation for the spectral density in terms of lambda+-}) we can hence conclude that
\begin{IEEEeqnarray}{rCl}\label{eq: Integral expression for the limiting spectral efficiency of the optimum receiver}
C^\opt(\snr,\beta,d) &=& \int_0^\infty \log_2(1+\snr \lambda)\rho(\lambda,\beta,d)\,\rmd \lambda \ . \quad
\end{IEEEeqnarray}
Explicit calculation of the integral finally yields \eqref{eq: Expression for C_opt with G-x-y-z}.
\end{IEEEproof}

\begin{rem}\label{rem: Remark on time-sharing for the optimum receiver}
Although \eqref{eq: Expression for C_opt with G-x-y-z} applies to $2\le d,\beta d\in\mathbb{N}^{+}<\infty$, total throughputs on the convex closure of the respective rates are achievable by means of time-sharing between different $(d,\beta d)$ points in the admissible set.
\end{rem}

We complete the  asymptotic analysis of the optimum receiver by means of extreme-SNR characterization.
Recall that a spectral efficiency $R$ is approximated in the low-SNR regime
 as $R\approx \frac{\calS_0}{3\mathrm{dB}}\left(\febno|_{\mathrm{dB}} -{{\febno}_{\min}}|_{\mathrm{dB}}\right)$, where $\calS_0$ denotes the low-SNR slope, ${\febno}_{\min}$ is the minimum $\febno$  that enables reliable communications, and $3\mathrm{dB}\triangleq 10\log_{10}2$ \cite{Shamai-Verdu-2001-fading}. The SNR and  $\febno$ are related via $\beta\snr = R \febno$. In the high-SNR regime the spectral efficiency is approximated as $R\approx \calS_\infty\left(\log_2\snr - \calL_\infty\right)$, where $\calS_\infty$ denotes the high-SNR slope (multiplexing gain), and $\calL_\infty$ denotes the high-SNR power offset \cite{Shamai-Verdu-2001-fading}.
The results are summarized in the following proposition (the proof is omitted due to space limitations).
\begin{prop}\label{prop: Extreme-SNR characterization of the optimum receiver}
Let $d$ and $\beta$ be as in Theorem
\ref{thm: Asymptotic Stieltjes transform and spectral density}.
Then, the low-SNR parameters of the optimum receiver read:
$\bigl(\tfrac{E_b}{N_0}\bigr)^\opt_{\min} = \ln 2$,  and $\calS^\opt_0 =\tfrac{2\beta d}{d(\beta+1)-1}$.
The high-SNR slope of the optimum receiver is given by
$\calS_\infty^\opt = \min(\beta,1)$,
while the high-SNR power offset satisfies
\begin{IEEEeqnarray*}{LR}
\calL_\infty^\opt = & \\
\begin{cases}
\bigl(\tfrac{1}{\beta}-1\bigr)\log_2(1-\beta) -  (d-1)\log_2\left(1-\tfrac{1}{d}\right)  , & \beta <1 \\
-(d-1)  \log_2\left(1-\tfrac{1}{d}\right) , &  \beta =1  \\
(\beta-1) \log_2(\beta-1)-\beta\log_2\beta& \\ \phantom{(\beta-1)} - (\beta d-1)\log_2\bigl(1-\tfrac{1}{\beta d}\bigr) \
 , & \beta >1 .
 \end{cases} & \quad \ \ \IEEEyesnumber
\end{IEEEeqnarray*}
\end{prop}
Comparison to RS-CDMA \cite{Shamai-Verdu-2001-fading} reveals that ${{\febno}_{\min}}$ and $\calS_\infty$ are identical in both settings. However, 
the low-SNR slope of regular sparse NOMA is strictly \emph{higher}, while the high-SNR power offset is strictly \emph{lower}, compared to
RS-CDMA. Regular sparse NOMA thus exhibits superior performance in extreme-SNR regimes.
The spectral efficiency coincides with that of RS-CDMA as $d\to\infty$ in \emph{all} SNR regimes, which is  also evident from \eqref{eq: Integral expression for the limiting spectral efficiency of the optimum receiver} and \eqref{eq: Final equation for the spectral density in terms of lambda+-} (cf.\ \cite{Shental-Zaidel-Shamai-ISIT-2017}).

%

\section{LMMSE Receiver}\label{sec: Linear MMSE Receiver}

In this section we turn to investigate the ergodic spectral efficiency of the LMMSE receiver. Recall that the corresponding error covariance matrix is given by $\Mat{M} = (\Mat{I}_K+\snr  \, \Mat{R})^{-1}$, where $\Mat{R}\triangleq \frac{1}{d}\Mat{A}^\dag\Mat{A}$ is the signature crosscorrelation matrix \cite{Verdu-Shamai-paper-3-99}. The signal-to-interference-plus-noise ratio (SINR) at the output of the receiver for user $k$ is $\frac{1}{M_{kk}}-1$, and  the spectral efficiency of the LMMSE receiver thus reads
\begin{IEEEeqnarray*}{rCl}
C^{\ms}(\snr,\beta,d)&=&\beta \, \bbE\bigl\{\tfrac{1}{K} \textstyle{\sum}_{k=1}^K \log_2\bigl(\tfrac{1}{M_{kk}}\bigr)\bigr\} \ . \IEEEyesnumber \label{eq: General expression for the SPEF of the LMMSE receiver}
\end{IEEEeqnarray*}
The following theorem characterizes the spectral efficiency of the LMMSE receiver in the large-system limit.

\begin{thm}
\label{thm: Asymptotic spectral efficiency of the LMMSE receiver}
Let the definitions and assumptions of Theorem~\ref{thm: Asymptotic spectral efficiency of the optimum receiver} hold. Then, the spectral efficiency of the LMMSE receiver converges as $N\to\infty$ to
\begin{IEEEeqnarray}{rCl}\label{eq: Final expression for the LMMSE SPEF - Overloaded regime - Form 1}
C^{\ms}(\snr,\beta,d)&=& \beta \textstyle{\log_2\left(\tfrac{1+\beta d \, \snr}{1+d\gamma \, \snr - \frac{d \EuS{F}(\gamma\snr,\tilde{\beta})}{4}}\right)} \ .
\end{IEEEeqnarray}
\end{thm}
\begin{IEEEproof}[Proof Outline]
Let ${\sfR}_{\Mat{R}}(z) \triangleq (\Mat{R}-z\Mat{I}_K)^{-1}$, $z\in\bbC^+$, denote the resolvent of $\Mat{R}$. Following \cite{Bordenave-Lelarge-2010}, it can be shown that the diagonal entries of ${\sfR}_{\Mat{R}}(z)$ converge in distribution to $\sqrt{\frac{d}{{z}}} X^b(\sqrt{dz})$ (cf.\
\eqref{eq: RDE for Xb}), which gives the Stieltjes transform of the limiting empirical eigenvalue distribution of $\Mat{R}$, $m_\Mat{R}(z)$. Applying analytic continuation we thus conclude that $M_{kk} \xrightarrow{d} \tfrac{1}{\snr} m_{\Mat{R}}(-\tfrac{1}{\snr})\triangleq M_1$.
Since the random variables $\{M_{kk}\}$ have a bounded strictly positive support, we may conclude by the continuity of $\log_2(\frac{1}{x})$ and uniform integrability of $\log_2(\frac{1}{M_{kk}})$, that $C^{\ms}(\snr,\beta,d) = \beta \log_2(M_1^{-1})$, leading to \eqref{eq: Final expression for the LMMSE SPEF - Overloaded regime - Form 1}.
\end{IEEEproof}
We note here that the time-sharing argument in Remark~\ref{rem: Remark on time-sharing for the optimum receiver} holds for the LMMSE receiver as well.
The extreme-SNR characterization of this receiver is given next.
\begin{prop}\label{prop: Extreme-SNR characterization of the LMMSE receiver}
Let $d$ and $\beta$ be as in Theorem
\ref{thm: Asymptotic Stieltjes transform and spectral density}. Then, the low-SNR parameters of the LMMSE  receiver read:
$\bigl(\tfrac{E_b}{N_0}\bigr)^\ms_{\min} = \ln 2$, and $\calS^\ms_0 =\tfrac{2\beta d}{(2\beta +1)d-2}$.
The high-SNR slope of the LMMSE receiver is given by
\begin{IEEEeqnarray}{rCl}
\calS_\infty^\ms =
\begin{cases}
\beta , & \beta <1 \\
\tfrac{1}{2}, & \beta=1\\
0, & \beta>1
\end{cases} \label{eq: high-SNR power offset of the LMMSE receiver}
\end{IEEEeqnarray}
while the high-SNR power offset satisfies
\begin{IEEEeqnarray}{rCl}\label{eq: LMMSE - Final expression for L_inf}
\calL^\ms_\infty &=&
\begin{cases}
 \log_2\left(\frac{1}{1-\beta}\right) + \log_2\left(\frac{d-1}{d}\right) \ , & \beta < 1\\
\log_2\left(\frac{d-1}{d}\right) \ , & \beta=1 .
\end{cases}
\end{IEEEeqnarray}
\end{prop}
Comparison to \cite{Shamai-Verdu-2001-fading} leads again to the same conclusions as drawn for the optimum receiver.


\section{Numerical Results}\label{sec: Numerical Results}

\begin{figure*}[t!]
    \centering
    \begin{subfigure}[t]{0.5\textwidth}
        \centering
        \includegraphics[width=\columnwidth]{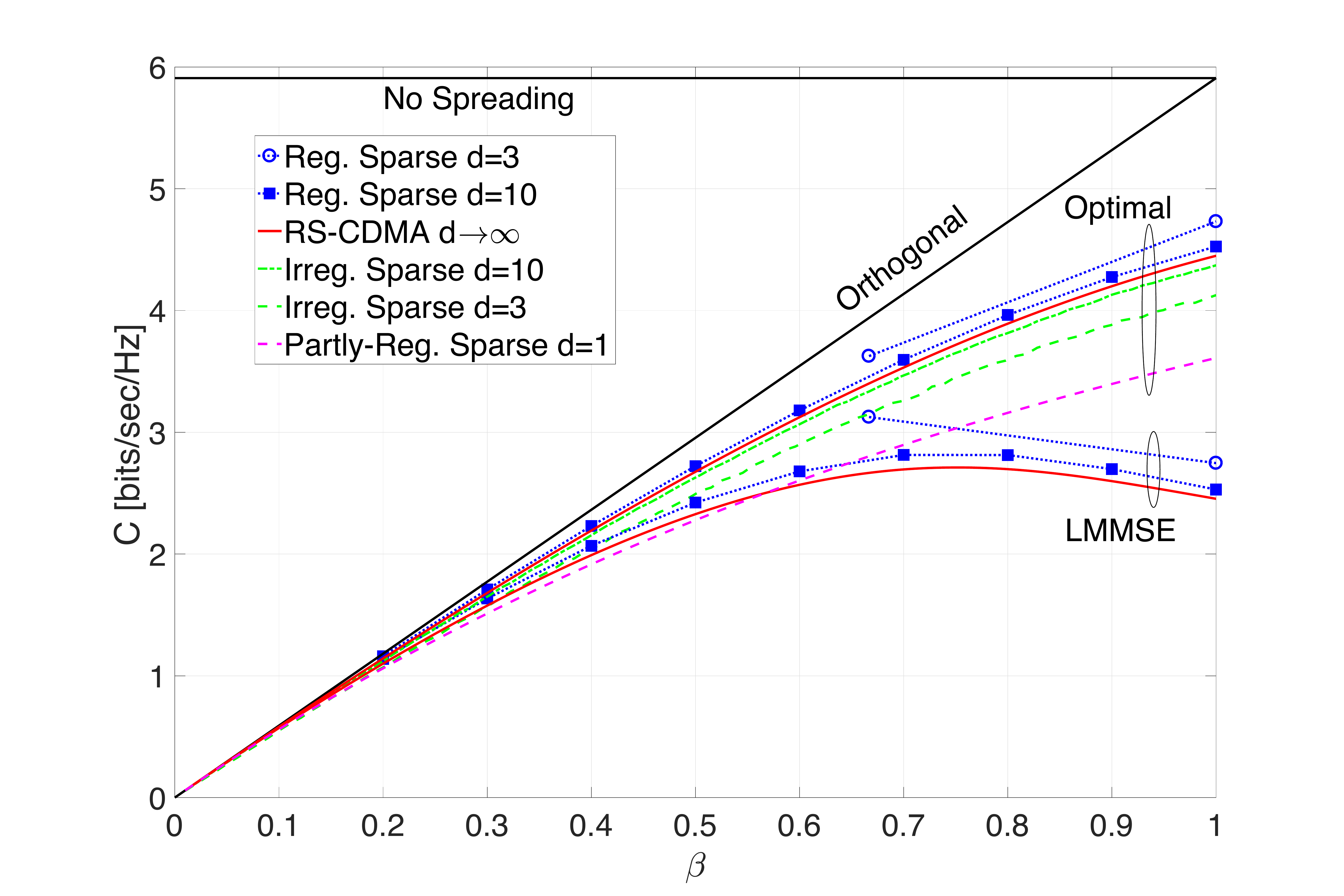}
        \caption{Underloaded regime}
    \end{subfigure}%
    ~
    \begin{subfigure}[t]{0.5\textwidth}
        \centering
        \includegraphics[width=\columnwidth]{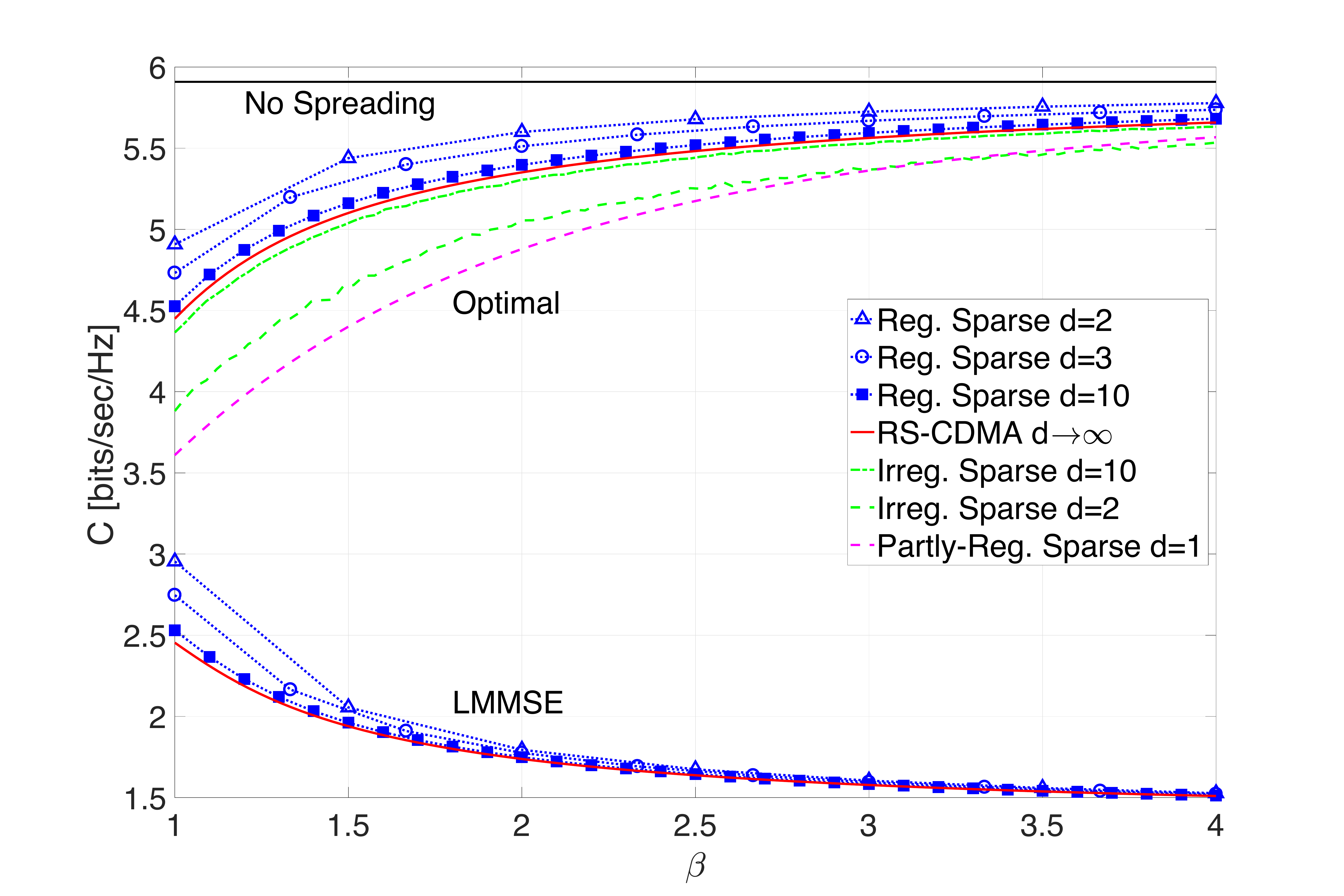}
        \caption{Overloaded regime}
    \end{subfigure}
    \caption{Limiting spectral efficiency vs.\ the system load $\beta$ for $E_{b}/N_{0}=10$ dB.}

    \label{fig: spef vs beta}
\end{figure*}

Some numerical results for the limiting spectral efficiency of regular sparse NOMA,
following \eqref{eq: Expression for C_opt with G-x-y-z} and \eqref{eq: Final expression for the LMMSE SPEF - Overloaded regime - Form 1},
are shown in Fig.~\ref{fig: spef vs beta}, and plotted as a function of the system load $\beta$ for fixed $\febno=10\textrm{dB}$.
The results were calculated for $d=2$ ($\beta\ge 1$), $3$ and $10$. The markers in the figure
designate the points for which $\beta d\in\bbN^+$.
The piecewise linear (dotted) lines
represent the achievable throughputs by \emph{exercising time-sharing} between integer $\beta d$ values (see Remark 4).
Fig.\ \ref{fig: spef vs beta} clearly indicates that regular sparse NOMA outperforms RS-CDMA \cite[Eqs.\ (9), (12)]{Verdu-Shamai-paper-3-99}. In the underloaded regime ($\beta <1$), the gap to the spectral efficiency of orthogonal transmissions \cite[Eq.\ (8)]{Verdu-Shamai-paper-3-99} is reduced. In the overloaded regime the scheme reduces the gap to the ultimate performance limit, as given by the Cover-Wyner bound \cite[Eq.\ (4)]{Verdu-Shamai-paper-3-99}, corresponding to the absence of spreading. Note also that the  spectral efficiency of regular sparse NOMA approaches that of RS-CDMA \emph{from above} with the increase of the sparsity parameter $d$, as already indicated in \cite{Shental-Zaidel-Shamai-ISIT-2017}. For the sake of comparison we also included in Fig.~\ref{fig: spef vs beta} the spectral efficiencies of irregular sparse NOMA \cite[Eqs.\ (28)--(32)]{Yoshida-Tanaka-ISIT-2006}, and partly-regular sparse NOMA
when the transmit energy is concentrated in a single orthogonal resource
(namely, $d=1$)
\cite[Eq.\ (13)]{Ferrante-Di-Benedetto-2015}\footnote{This seems to be the only case where the spectral efficiency is \emph{fully} analytically tractable in this framework (cf.\ \cite{Ferrante-Di-Benedetto-2015}).}.
%
The results
reinforce
the initial observations made in  \cite{Shental-Zaidel-Shamai-ISIT-2017}, indicating that irregular sparse NOMA schemes exhibit degraded performance,
due~to~the fact
that some resources may be left unused (and in the fully irregular case some users may end up without any designated resources).
%
Irregular sparse NOMA schemes are also observed~to approach the spectral efficiency of RS-CDMA with the increase of $d$, however the approach to the limit is from \emph{below} (cf.\ \cite{Ferrante-Di-Benedetto-2015}).



\section{Concluding Remarks}\label{sec: Concluding Remarks}

Regular sparse NOMA has been investigated in this paper in the large-system limit. Considering the optimum receiver (whose performance can be approached using \emph{practical} MPAs even in \emph{overloaded} regimes), and the simple LMMSE receiver, the respective spectral efficiencies were  expressed in \emph{closed explicit form}, and shown to outperform the achievable throughputs of both RS-CDMA and irregular sparse NOMA. It is crucial to  emphasize that the underlying NOMA system model is markedly different here from previously analyzed settings.
Firstly, the random matrices are sparse, as opposed to standard RS-CDMA~\cite{Verdu-Shamai-paper-3-99}, where performance is governed by the \emph{Mar\v{c}enko-Pastur law}.
Secondly, the entries of the signature matrix are not i.i.d., as opposed to Poissonian irregular sparse NOMA (\eg, \cite{Yoshida-Tanaka-ISIT-2006}). The setting also differs from the recently analyzed partly-regular scheme \cite{Ferrante-Di-Benedetto-2015} (time-hopping CDMA), where sparse random spreading sequences with a fixed number of non-zero entries per time frame are employed. This is since in the current setting the number of non-zero entries is identical and fixed also in each \emph{row} of the signature matrix, in contrast to \cite{Ferrante-Di-Benedetto-2015}. Finally, the setting differs also from the sparse models considered, \eg, in~\cite{Guo-Baron-Shamai-Allerton-2009}, where the limiting average sparsity amounts to a fixed (small) \emph{fraction} of the dimensions (implying linear scaling); and \cite{Guo-Wang-2008}, where the number of non-zero signature entries amounts to a vanishing fraction of its dimension, but is still \emph{infinite} in the large-system limit. We conclude by noting that our observations facilitate the understanding of the potential performance gains of sparse NOMA, and advocate employing \emph{regular} schemes as a key practical tool for enhancing performance of future highly loaded cellular systems.




\section*{Acknowledgment}
The work of B.\ M.\ Zaidel and S.\ Shamai (Shitz) was supported 
by the Heron consortium via the Israel Ministry of Economy
and Industry.





\end{document}